\documentclass[runningheads]{llncs}

\usepackage{graphicx}
\usepackage{booktabs}
\usepackage{mathtools}
\usepackage{amsmath}
\usepackage{amsfonts}
\usepackage{hyperref}
\usepackage[colorinlistoftodos]{todonotes}
\usepackage{pdfpages}
\usepackage{commath}
\usepackage{tabularx}
\usepackage{url}
\usepackage[T1]{fontenc}
\usepackage{subfigure}
\usepackage{floatrow}
\usepackage{color}

\usepackage[title]{appendix}

\newfloatcommand{capbtabbox}{table}[][\FBwidth]

\DeclareCaptionLabelFormat{andtable}{#1~#2  \&  \tablename~\thetable}

\usepackage{xspace} 
\newcommand{\cnc}[0]{\textsc{C\&C}\xspace}
\newcommand{\system}[0]{MORTON\xspace}
\newcommand{\etal}[0]{\textit{et al.}\xspace}
\newcommand{\baywatch}[0]{Baywatch\xspace}

\begin{document}
	
\title{\system: Detection of Malicious Routines in Large-Scale DNS Traffic}


\authorrunning{Daihes et al.}
\author{Yael Daihes\inst{1,2} \and
Hen Tzaban\inst{2} \and
Asaf Nadler\inst{1,2} \and
Asaf Shabtai\inst{1}}

\institute{Software and Information Systems Eng., Ben-Gurion University of the Negev \and 
Akamai Technologies Inc. \\
}

\maketitle

\begin{abstract}
In this paper, we present \system, a method that identifies compromised devices in enterprise networks based on the existence of routine DNS communication between devices and disreputable host names.
With its compact representation of the input data and use of efficient signal processing and a neural network for classification, \system is designed to be accurate, robust, and scalable.
We evaluate \system using a large dataset of corporate DNS logs and compare it with two recently proposed beaconing detection methods aimed at detecting malware communication.
The results demonstrate that while \system's accuracy in a synthetic experiment is comparable to that of the other methods, it outperforms those methods in terms of its ability to detect sophisticated bot communication techniques, such as multistage channels, as well as in its robustness and efficiency.
In a real-world evaluation, which includes previously unreported threats, \system and the two compared methods were deployed to monitor the (unlabeled) DNS traffic of two global enterprises for a week-long period; this evaluation demonstrates the effectiveness of \system in real-world scenarios and showcases its superiority in terms of true and false positive rates.

\keywords{DNS, PSD, Neural Networks, Botnet}

\end{abstract}


\section{\label{sec:introduction}Introduction}
Enterprise networks are prominent targets of malware~\cite{kotzias2019mind}. 
After a malware has been downloaded and executed on an enterprise device, the device effectively becomes part of a botnet controlled by a remote attacker.
The attacker exchanges control messages with its bots through a command and control (\cnc) channel, by which the attacker instructs the bots to perform attacks, such as data theft~\cite{nadler2019detection} and DDoS attacks~\cite{welzel2014measuring}.
To protect against these attacks, numerous studies have focused on the detection of bots' \cnc communication, for example, by processing the domain name system (DNS) protocol logs~\cite{zhauniarovich2018survey}.

The DNS protocol is a core component of the Internet whose primary goal is to translate a host (domain) name to its IP address. 
Bots often rely on the DNS protocol to acquire the IP address of their \cnc server, because the IP address can be changed whenever the existing host becomes blocked or suspected.
Modern botnets frequently change their C\&C servers' IP addresses and domain names using techniques, such as fast flux~\cite{holz2008measuring,schales2015fcce} and domain generation algorithms (DGA)~\cite{plohmann2016comprehensive}, in order to avoid being detected.
The fact that in many cases the botnet utilizes the DNS protocol has led to extensive research on the detection of malicious domain names in DNS logs, by tracking the patterns of fast flux and DGA techniques and identifying devices that query these domain names~\cite{zhauniarovich2018survey}.
Despite the fact that the practice of classifying bots based on the DNS protocol through malicious domain name detection has been widely adopted, it fails to detect all botnet communication, and less than 20\% of all malicious domain names are reported and eventually added to DNS blocklists~\cite{kuhrer2014paint}.
Given this, the goal of this study is to explore and evaluate methods for classifying the ~\emph{devices} identified as bots in large-scale enterprise DNS traffic, an independent approach that complements malicious domain name detection. 
More specifically, we are interested in developing a method to detect bot devices if they engage in routine communication to disreputable host names.

Prior studies~\cite{hu2016baywatch,shalaginov2016malware} that attempted to identify bots without relying on malicious domain name detection focused on the detection of a bot communication technique called ``beaconing'' --- a communication technique in which a recurring message is sent from the bots to predetermined domain names so that the attacker can know the latest time at which a bot was available.
\emph{Connection pairs (CPs)}, i.e., the outgoing DNS queries from a \emph{specific device} to a \emph{specific domain name}, that exhibit periodicity are suspected as botnet beaconing communication which should be further investigated.

Existing beaconing detection methods 
suffer from a couple of limitations.
First, the number of CPs in large-scale enterprises is extremely large.
Therefore, to keep up with the scale, these methods must either filter the CPs with less suspicious host names prior to classification to provide acceptable performance, and/or, must be executed less frequently, which will lead to longer detection times and delayed detection.
Second, attackers may try and evade the above-mentioned periodicity detection methods by communicating through multiple host names, which, despite being periodic, are not captured within any single CP.
One notable example of a bot communication technique that involves multiple host names is multistage channels~\cite{MITREMatrix} (MSC), in which a bot communicates with a set of host names throughout its lifecycle; for instance, the TrickBot and Emotet banking Trojans, which are considered major business security threats ~\cite{rendell2019understanding}, make use of MSCs, thus demonstrating the importance of identifying bot communication that involves multiple host names.
Further details on bot communication that involve multiple host names is provided in appendix~\ref{sec:appen1}.

In this paper, we present \textbf{\system}: a method for \textbf{m}alici\textbf{o}us \textbf{r}outine de\textbf{t}ecti\textbf{on} on large-scale DNS traffic.
\system analyzes the DNS communication logs of large enterprise networks and identifies devices that periodically communicate with a set of disreputable domains.
Compared to prior studies, \system is more efficient and thus detects threats sooner than other proposed methods, enabling organizations to block these threats faster and prevent the attackers from achieving their attack goals against the organization.
Specifically, \system classifies the outgoing DNS requests made by a device directly, rather than indirectly classifying CPs, which decreases the input size and eliminates the need for heavy filtering to reduce the number of CPs examined.
Direct classification of devices as bots has two main advantages.
First, direct classification is faster, because it involves fewer classification tasks (i.e., the number of enterprise devices is significantly smaller than the number of CPs).
Second, direct classification can be used to detect malicious routines that include multiple host names, such as MSC and multihop attacks.

The evaluation presented in this study compares \system with two recently proposed methods designed to identify malware beaconing: \baywatch~\cite{hu2016baywatch} and the adaption of WARP~\cite{elfeky2005warp} to DNS logs, as proposed by Shalaginov \etal~\cite{shalaginov2016malware}.
Two experiments were performed to compare the performance of the three methods on two tasks: a labeled and an unlabeled task.
For the first task, we used a dataset consisting of \textit{real} DNS logs that were collected from 16,000 managed devices of eight worldwide corporate networks over the course of a week. 
The devices in the dataset are assumed not to be infected with bots (due to the scarcity of \cnc communication on managed enterprise networks).
To obtain malicious labels (i.e., bots), we selected a subset of the \emph{benign} devices in the dataset into which we \emph{inject} DNS queries of malware beaconing and MSC communication, mimicking the behavior of known malware, such as CobaltStrike~\cite{CobaltStrike} and Empire~\cite{EmpireMalware}.
The labeled dataset generated, which we refer to as the \emph{synthetic dataset}, is used to evaluate three important aspects of bot detection methods: accuracy, robustness to noise, and scalability.

Our results demonstrate that malware beaconing is successfully detected by all of the evaluated methods ($AUC>0.85$), however MSCs are detected by \system ($AUC>0.85$) but overlooked by the other two methods ($AUC<0.72$).
We measure robustness using a score that embodies the detection rate as a function of unobserved traffic (i.e., noise).
For beaconing detection, there is a tie, with both \system and Baywatch achieving the highest robustness score, but for MSCs, \system dominates the other methods.
Moreover, we found that it only took \system 70 seconds to be applied on the entire test set (containing 700,000 CPs), a rate which was almost a hundred times faster than that of Baywatch and thousand times faster(!) than WARP.

For the second task of evaluating a real-world unlabeled scenario, \system and the two compared methods were deployed to monitor two large-scale enterprise networks for a one-week period.
The detections made by the three methods during this monitoring period were validated using VirusTotal. 
\system achieved a true positive rate of 62\% which was almost three times more than the other two methods. To our understanding, \system dominates in terms of the true positive rate due to its design which is not only aimed at detecting periodic behavior, but rather \emph{malicous} periodic behavior. A few case studies were analyzed to demonstrate the practicality of \system in a real-world, unsupervised environment and to understand the reason behind the performance difference between the methods. 
Furthermore, we provide a qualitative analysis of selected detections made by the algorithms to demonstrate the importance of periodicity detection in general, and MSC detection in particular.

We summarize our paper’s contributions as follows:
\begin{enumerate}
    \item We introduce \system, an accurate, robust, and efficient method for detecting routine bot communication in large-scale DNS traffic.
    \system is capable of detecting bot communication techniques that have thus far been overlooked by other periodicity detection methods (e.g., MSC). 
    \item We provide a comparison of \system and two recently proposed methods on a large-scale DNS labeled dataset to evaluate accuracy, robustness, and efficiency.
    \item We present the results of \system in comparison to two recently proposed periodicity detection methods on real-world, large-scale DNS unlabeled logs that were produced by two worldwide enterprises over the course of one week. \system's true positive rate supersedes that of the other two methods by as much as three times.
    Several case studies were explored to demonstrate the practicality of \system in a real-world, unsupervised environment and help explain why \system performs far better than the other methods.
\end{enumerate}

\section{Related Work} \label{sec:short related works}
The detection of malicious activity in DNS traffic has been thoroughly studied over the last decade. 
To identify the malicious activity, relevant studies targeted the detection of enterprise devices, external hosts names, or their combination (CPs).
The majority of studies targeted just external host names by identifying hosts involved in fast flux networks~\cite{holz2008measuring}, data exfiltration~\cite{nadler2019detection}, or DGAs~\cite{plohmann2016comprehensive}.
These studies are effective in identifying malicious hosts~\cite{zhauniarovich2018survey}, but they are limited to specific use cases that do not adequately cover the threat landscape.

\vspace{-4mm}
\begin{table} 
\scriptsize
\newcolumntype{b}{X}
\newcolumntype{s}{>{\hsize=.5\hsize}X}
\begin{tabularx}{\textwidth}{|s|s|b|} 
\hline
\textbf{Studies} & \textbf{Targets} & \textbf{Conditions} \\ \hline
\system & Devices & Periodic communication to disreputable hosts  \\ \hline
  ~\cite{yeh2018malware,jiang2019new,haffey2018modeling,tran2015host,hubballi2013flowsummary,shalaginov2016malware}  & CPs  &  Periodic communication to disreputable hosts  \\ \hline
 ~\cite{manasrah2020botnet,yu2010online,huynh2019frequency,hu2016baywatch} & CPs & + Infections of multiple devices  \\ \hline
 ~\cite{sivakorn2019countering,bilge2012disclosure,oprea2018made,khan2019adaptive} & Devices or Hosts & + Non-DNS logs (e.g., NetFlow) \\ \hline
~\cite{sato2012extending,gao2013empirical,zhauniarovich2018survey,plohmann2016comprehensive,holz2008measuring,nadler2019detection} & Hosts & DGA, Fast flux or Data exfiltration  \\ \hline
\end{tabularx}
\caption{Summary of related research on malicious activity detection using DNS traffic.}
\label{tab:related_work_tab}
\end{table}

\vspace{-15pt}
To compliment host-targeting studies, another category of studies focus on devices that engage in periodic communication with disreputable hosts (i.e., device detection).
Device detection studies are often concerned with the potential inaccuracy of malicious activity detection based \emph{only} on periodic communication.
To prevent such inaccuracy, these studies have employed one of the following techniques: (1) using additional context (other than DNS logs) ~\cite{sivakorn2019countering,bilge2012disclosure,oprea2018made,khan2019adaptive}, (2) assuming similarity between infected devices~\cite{manasrah2020botnet,yu2010online,huynh2019frequency,hu2016baywatch}, and/or (3) classifying combinations of devices and hosts (i.e., CPs) rather than devices~\cite{yeh2018malware,jiang2019new,haffey2018modeling,tran2015host,hubballi2013flowsummary,shalaginov2016malware}.
A summary of the studies appears in Table~\ref{tab:related_work_tab}.

The drawbacks of the abovementioned techniques, as shown in this study, are decreased effectiveness against bot communication that involves multiple external host names (e.g., MSC), reduced efficiency on large-scale DNS traffic, and limited use to only when the additional context is provided.
\system addresses these drawbacks by directly classifying the outgoing DNS queries made by a device rather than performing indirect classification of connection pairs; this is accomplished by combining signal processing with a neural network to efficiently and accurately classify compromised devices.

\section{\system} \label{sec:system} 
\subsection{Overview}
\begin{figure*}[ht]
\centering
\includegraphics[width=1.0\textwidth]{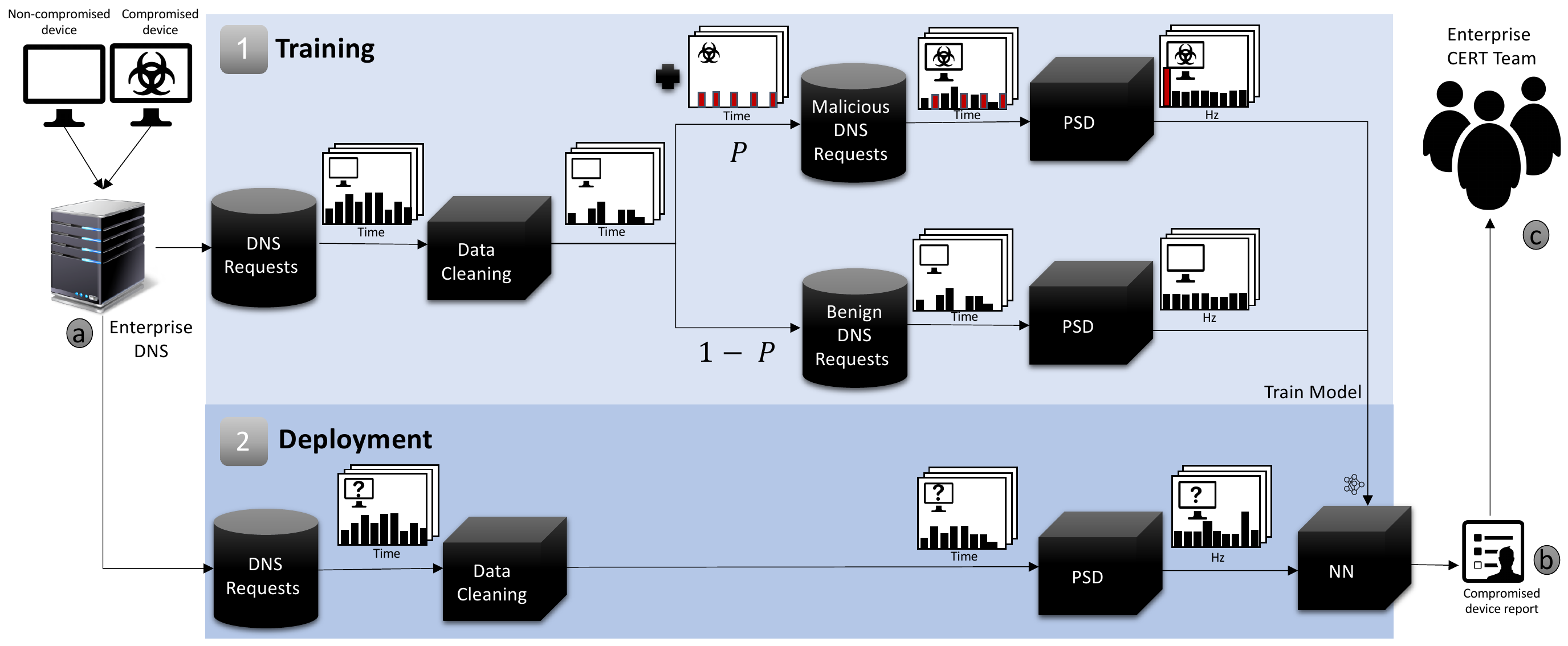}
\caption{\system training and deployment. Training (1) includes legitimate DNS traffic which is divided into the traffic of benign devices and the traffic of devices into which malicious bot communication was injected, thereby creating the synthetic data. The synthetic labeled data is used to train a neural network that classifies a newly observed time series of DNS requests, based on whether the requests were made by a bot or not. In deployment (2) a neural network processes and classifies new DNS requests series. 
In turn, (a) \system identifies new bots, and (b) reports them for further investigation by an enterprise's CERT team (c).}
\label{fig:scheme}
\end{figure*}

\system consists of two phases: data processing and classification (see Figure~\ref{fig:scheme}).
The data processing phase transforms the input, a time series of outgoing DNS queries, to a power spectral density (PSD) vector that characterizes the intensity of periodic communication at various frequencies.
In this phase, we first filter the time series of DNS queries from reputable host names that are not likely to be involved in malicious activity.
Then, the filtered time series of DNS queries are aggregated by time.  
The data processing ends by applying discrete Fourier transform (DFT) on the aggregated DNS queries to produce the PSD vector.
In the classification phase, a neural network model classifies the PSD vector based on whether the device that made the DNS queries is a bot or not.

\vspace{-4mm}

\vspace{-10mm}
 
\subsection{\label{subsec:definitions}Definitions} 
An outgoing DNS query ($Q$) made by internal device ($D$) to a host name $H$ at time ($T$) is represented as a tuple $Q=<D, H, T> $.

The input to \system is a series of DNS queries made by a specific device ($D_i$) within a predetermined timeframe that starts at $T_s$ and ends at $T_e$.
The input is formally defined as $I_{(D_{i}, T_{s}, T_{e})} = \{ Q | (D = D_{i}) \wedge (T_{s} \leq T < T_{e}) \}$.

\subsection{\label{subsec:data-processing}Data processing} 
\textbf{Filtering:} \system's input might contain queries to host names that are assumed to be trusted.
Accordingly, the goal of data filtering is to remove queries to reputable host names from the input to improve the accuracy and decrease the processing time.
These benefits make data filtering processes very common in methods that attempt to detect malicious routines~\cite{hu2016baywatch,shalaginov2016malware}.

The data filtering process applied in \system is based on the ideas of global and local reputation ranking as described in~\cite{hu2016baywatch}.
Global reputation dictates that a host name is reputable and therefore should be trusted if it is listed in publicly available lists that are often associated with benign host names (e.g., the Alexa Top 1M list).
Local reputation complements global reputation by dictating that the host names are also regarded as trusted if they are queried by a sufficiently large portion of devices in the network.

We define a \emph{filtering function} $C$ that outputs a true value indicating if a $Q$ should remain.
The filtering function respects both the global reputation and the local reputation, using special parameters, $\phi_{G}$ and $\phi_{L}$.
$\phi_{G}$ defines the number of highest-ranking Alexa top 1M domains that are regarded as trusted.
$\phi_{L}$ defines the minimal rate of querying devices to consider a host name as trusted.
For instance, if $\phi_{G}$ is set at 100,000 and $\phi_{L}$ is set at $0.3$, then host names that are ranked below 100,000 in the Alexa top 1M list or are accessed by at least 30\% of the devices within the examined timeframe are regarded as trusted and are filtered from the input. Formally, the filtering is applied to the input using the filtering function: $I^{c}_{(D_{i}, T_{s}, T_{e})} = \{ Q | (D = D_{i}) \wedge                                        (T_{s} \leq T < T_{e})  \wedge (C(Q) \text{ is True}) \}$.


\noindent\textbf{DNS Query Aggregation:} The DNS query aggregation stage transforms a filtered input, which is a series of outgoing DNS queries, into an aggregate series of DNS queries, counting the number of DNS queries made by a device within each timeframe.
Aggregating the original signal inherently consists of a loss of information, but it has several important advantages.
First, the resulting aggregation is dramatically smaller than the filtered input, thereby reducing the required processing time of such data.
Second, aggregating the data generates new input of a \emph{constant} size which is the number of timeframes examined, \emph{regardless} of the filtering function configuration, thus allowing \system to observe the signal without information-less padding (usually used for varied sized inputs).
Lastly, \emph{aware} attackers can design their bots so they jitter (i.e., wait for a random period of time before every communication) to appear less periodic and evade detection.
Our use of aggregating results to discrete timeframes (e.g., hourly query counts) limits an aware attacker's ability to evade our detection.
For instance, an aware attacker that knows \system was configured to use hourly timeframes would have to choose between communicating within an hour of the communication time and becoming detected or delaying communication by over an hour thus reducing the botnet's effectiveness.
Formally, $T_{(D_{i}, T_{s}, T_{e})}=(t_{1},t_{2},t_{3},..,t_{i},..,t_{N})$ defines the aggregation of the filtered input, where $t_i$ represents the number of overall DNS queries that were made in the $i$-th timeframe, and $N$ is the overall number of $\lambda$ seconds timeframes.


\system can be configured to use different values of $N$ and $\lambda$. 
Lower $\lambda$ values will result in shorter timeframes that provide better data granularity for accuracy, while higher values will result in longer timeframes that provide greater efficiency through a smaller representation.
Higher $N$ values will result in the method inspecting more timeframes, thus making it more sensitive to detections that occur over a period of time, but less efficient, because the input becomes larger.
For brevity, for the remainder of the paper we refer to the DNS query aggregation $T_{(D_{i}, T_{s}, T_{e})}$ simply as $T$.

\noindent\textbf{PSD Computation:} A power spectral density (PSD) vector is a common data representation form for periodicity detection tasks.
The PSD vector is defined as the squared magnitude of the discrete Fourier transform (DFT) coefficients for a given input.
PSD vectors are used to estimate the spectral density within a time series signal, which later allows classifiers to more easily identify periods in which a repetitive behavior takes place.
Intuitively, a PSD vector can be thought of as the ``intensity'' of the rate of events at particular time frequencies. 
Accordingly, a frequency entry within the PSD vector that has a high value indicates a routine event occurring within that frequency. 

First, DFT is applied to the DNS query aggregation $T$ as shown in Equation~\ref{eq:dft}:
\begin{equation} \label{eq:dft}
DFT(T,k)=\sum_{n=0}^{N-1} t_n^{e^{-i 2\pi}\frac{kn}{N}}
\vspace{-4mm}
\end{equation}
where $k=0,1,...,N-1$.

By definition, every PSD vector entry (frequency) is defined as described in Equation~\ref{eq:psd_comp}:
\begin{equation} \label{eq:psd_comp}
F_i = \norm{DFT\Big(T,0\Big)}^{2}
\end{equation}
resulting in a PSD vector of the form~\ref{eq:psd}:
\begin{equation} \label{eq:psd}
PSD(T) = \left( F_0, F_1, ..., F_{\frac{N-1}{2}-1} \right)
\end{equation}

For example, if the number of timeframes is $N=168$, then the output PSD will have $\frac{N-1}{2}=83$ entries, as shown below:
\[
(F_0 = 0.145, F_1=0.03, .., F_{82}=0.01)
\]

\noindent\textbf{Normalization:} The values of the PSD vectors are unbounded and of different magnitudes, and therefore they must be normalized prior to classification.
The normalization of PSD vectors in \system scales the amplitude of every frequency to a value between zero and one, which is proportional to the original values that appear in the training set.
The resulting normalized PSD vector is guaranteed to maintain the same ``distribution of energy'' for every frequency and a consistent scale for classification.
The normalized PSD vector is henceforth referred to as $\overline{PSD(T)}$.

\noindent\textbf{Classification:} \system trains a vanilla feedforward neural network (NN); this type of NN is commonly used for classification tasks with well structured input due to its performance.
The NN parameters ($\theta$) are learned in training, and the network is applied as a function ($F$) on the normalized PSD to output a classification result ($Y$); 
formally, $ Y = F(\theta, \overline{PSD(T)})$.
The result ($Y$) is a continuous value ranging from zero to one, where higher values indicate greater certainty that the input DNS queries were made by a bot.

\section{Labeled Evaluation} \label{sec:eval}
The environment used to conduct this evaluation is described in Appendix~\ref{sec:appen2}.
\subsection{\label{subsec:dataset}Dataset}

To generate labeled data we used a dataset consisting of \textit{real} DNS logs from Akamai's DNS traffic that cover eight enterprise networks in a variety of time zones. 
The DNS logs describe queries made by 16,000 devices to 1,262,527 host names.
The number of unique connection pairs (i.e., device to host name) is 1,893,221.
Every DNS log line contains the DNS query timestamp, the device identifier, and the queried host names, as described in Section~\ref{subsec:definitions}.
We assume that managed enterprise devices are rarely part of a botnet. 
Therefore, we label all of the devices in the dataset as benign, except for a random subset of devices (5\%) into which we inject synthetic malicious bot communication traffic that mimics the behavior of known malware, such as CobaltStrike~\cite{CobaltStrike} and Empire~\cite{EmpireMalware}. 
The 16,000 devices are split into a training set of 10,000 devices and a test set of 6,000, while maintaining the proportion of labels.

\subsection{Parameters}

The malicious bot communication traffic injected in order to generate our synthetic training data is configured using three variables: (1) the time interval (in minutes) between consecutive malicious queries, (2) the number of DNS queries that are sent in every interval, and (3) the malicious host name (or names) to which the DNS queries are sent.

The inter-arrival time between consecutive malware beaconing varies based on the attacker. 
To best mimic typical attackers, we rely on public frameworks such as CobaltStrike ~\cite{CobaltStrike} and Empire ~\cite{EmpireMalware}.
These frameworks enable attackers to configure parameters such as: the inter-arrival time between beacons in minutes/hours, the contacted hosts, number of queries made, and the jitter limit (limit of random period of time added before every communication, which defines the level of deviation from true periodicity). 
Accordingly, in our configuration, the inter-arrival time is an integer number, and it is sampled uniformly, with possible values ranging from 120 minutes (i.e., communicate every two hours) to 720 minutes (i.e., communicate every 12 hours), as allowed by abovementioned frameworks. 
The number of queries made for each interval is also sampled uniformly, with values ranging from five to 15. 

In order to account for noise such as: network delays, packet drops and intentional jitter, \system aggregates DNS queries to predetermined time windows, therefore canceling out noises smaller than the time window.
Within the scope of this evaluation, we define the time window size to be exactly one hour, to cancel noise of up to 30 minutes; less than excepted by network delays, drops and jitters. 
We set the destination to either single or multiple hosts, so we can evaluate both scenarios. The host names used are names that did not appear in the dataset. For the MSC scenario, we select the number of host names to be used (between three and six), and for every DNS query, we select random host names from the host names set.

Dataset filtering is performed using the following configuration: $\phi_{G} = 500,000$ and $\phi_{L} = 0.03$, i.e., only the top 500,000 host names on the Alexa Top 1M list and host names that are queried by at least 3\% of the devices throughout the past week are omitted from the dataset.
The filtering step supports the notion that not many devices within the same enterprise are likely to be infected with the same threat. By sticking to that notion, the filtering eliminates a common false positive within the periodicity detection realm,  domains used for intentional enterprise software updates. 

\subsection{Methods compared}
The experiment compares the following methods: \system, \baywatch~\cite{hu2016baywatch}, and WARP~\cite{shalaginov2016malware}, on the task of detecting beaconing and MSC.
The \baywatch method is evaluated using two different settings, namely, a fast setting and an accurate setting, to account for the range of achievable results by this method.
Our comparison of the methods focuses on the accuracy of detection, robustness to noise in the data, and the run-time performance.

\textbf{\system:}
\system performed best with a neural network architecture consisting of three hidden layers with 25, 55, and 25 neurons respectively, the ReLU activation function, and an output layer with the sigmoid activation function.
Our evaluation and the results presented are based on this architecture.

The training of the neural network is conducted with a dropout setting on the hidden layers with a rate of 0.1; this is done to regulate the training, so it will use all of the neurons and achieve its optimal performance. 
Early stopping is defined so that if the validation loss remains roughly unchanged for five consecutive epochs, the training stops.
The Adam optimizer is used to minimize the binary cross-entropy loss. 

\textbf{Baywatch:} The \baywatch method has exactly two parameters, $C$ and $m$, both of which are explained below, because they significantly affect the trade-off between the run-time of the method and the desired level of confidence.
The $C$ parameter is a continuous value between zero and one, and it is used to indicate the level of certainty. 
The $m$ parameter is a discrete positive value that is used to construct a baseline of $m$ non-periodic signals, based on which the method is able to establish the significance of the periodicity signal observed.
For every CP between a device and a queried domain name, \baywatch generates $m$ random permutations of the input time series.
Then, if the maximal value of the PSD belonging to the original time series is within the $C$ percentile of the maximal value of the PSD belonging to the $m$ random permutations of the time series, the CP of the machine and queried domain name is classified as a pair that is engaged in a periodic communication channel.
Based on this behavior, assignments of higher $m$ values would produce a more robust baseline to assure more accurate results while increasing the run-time of the overall autocorrelation function (ACF) linearly.
Accordingly, the evaluation includes two different settings to best reflect Baywatch's performance: an \emph{accurate setting} ($m=100$), as was evaluated in the original study, and a \emph{fast setting} ($m=10$). 
For each of these settings, we evaluate different confidence values that are commonly used for a low false alarm rate in security systems, i.e., $C=0.9, 0.99, 0.999$. 
For the sake of brevity, we henceforth refer to the two settings as \baywatch-10 and \baywatch-100.

\textbf{WARP:} Shalaginov \etal presented a general method for periodicity detection. 
Within the general method, CPs of internal devices and external host names are extracted and processed to return the length of the minimal periodic cycle.
Processing specific CPs consists of computing the time difference between consecutive DNS queries, smoothing the time difference, and replacing the smoothed time difference values with unique symbols that form a string.
The string formed is provided to an underlying periodicity detection algorithm that processes strings.
Shalaginov \etal proposed four potential underlying periodicity detection algorithms, and our implementation relies on one of them, namely, ``WARP''~\cite{elfeky2005warp}.
The other periodicity detection algorithms that were mentioned in the study require parameter settings that were not provided in the paper; WARP is the only algorithm that we could guarantee would perform similarly to the version evaluated by Shalaginov \etal
Our beaconing and MSC experiments consist of a repeating query with a constant inter-arrival time.
Therefore, if the returned minimal periodic cycle is one, we classify the CP as periodic, because all time differences between consecutive queries are equal after smoothing.
Otherwise, we classify the CP as non-periodic.
The smoothing variance is defined by a parameter $s$, so that the modulo of the time difference from $s$ is removed.
Accordingly, higher $s$ values make the classification more sensitive (i.e., having a higher recall score but a lower precision score).

\subsection{Evaluation results}
\begin{figure*}[ht]
\centering
\includegraphics[width=1.0\textwidth]{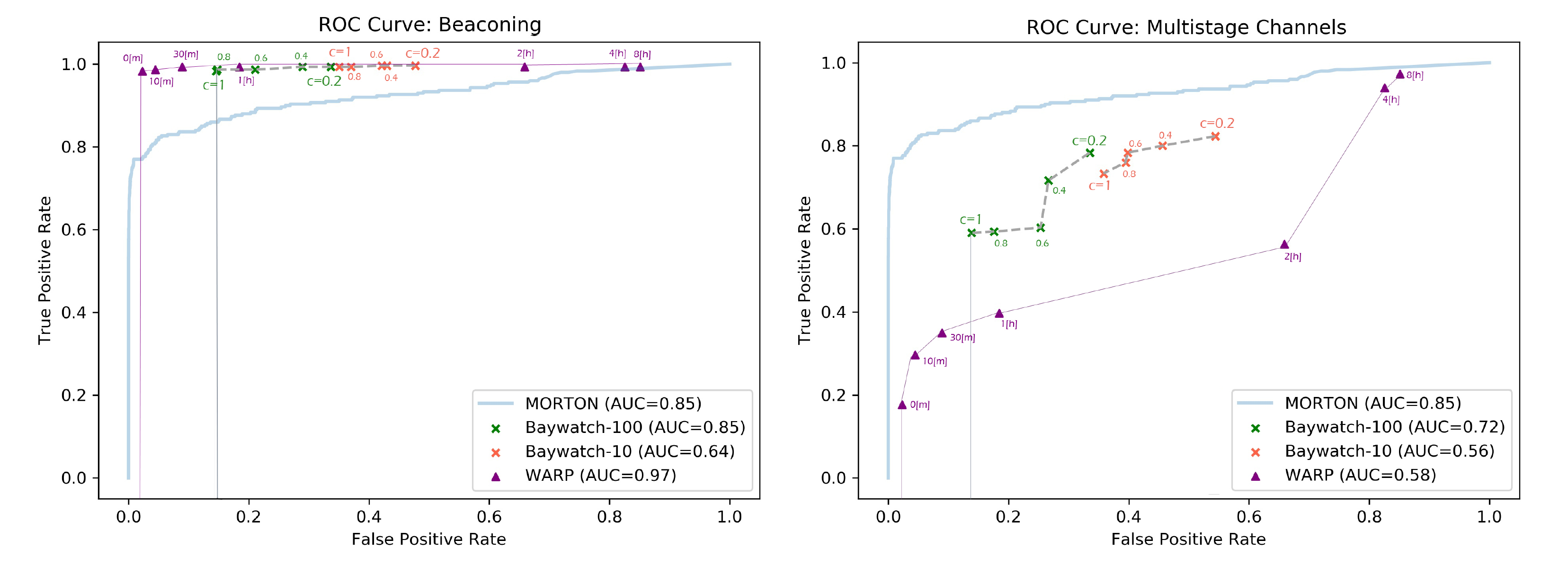}
\caption{Receiver operating characteristic curve
\label{fig:roc-sssd}
 of \system, \baywatch, and WARP, with different $m$ values, on the test set.}
\end{figure*}
\textbf{Accuracy:}
Our first experiment deals with the accuracy of the evaluated methods by analyzing the receiver operating characteristic (ROC) curves that describe the methods' detection rates on the test set.
The ROC curves for detecting beaconing and MSC are presented in Figure~\ref{fig:roc-sssd}.
The graphs are analyzed from two perspectives: the area under the curve (AUC) and the TPR for a fixed and low FPR (1\%), which is often the setting in automated cybersecurity systems, in order to reduce the number of false alarms that must be investigated by an enterprise CERT.

For beaconing detection, WARP is the most accurate method (AUC=0.97), followed by \system (AUC=0.85), Baywatch-100 (AUC=0.85), and finally by the fast Baywatch-10 (AUC=0.64).
The AUC scores show that all methods are capable of detecting beaconing with high accuracy, with WARP performing best.
Despite that, for a low FPR value of 1\%, only \system was able to produce adequate results, with a TPR value of 77\%, compared to the other methods that were unable to produce any detection with the low FPR of 1\%.
For \baywatch, setting the confidence parameter $C$ to 1.0 resulted in a minimal FPR of 14.4\% thus reflecting \baywatch's lack of sensitivity, which is addressed by the authors who mentioned the need to apply additional methods to reduce the FPR further.
For WARP, the lowest $s$ value of zero (i.e., highest confidence) results in a minimal FPR of 3\%, which is more sensitive than \baywatch but less sensitive than \system.
Effectively, the automated use of these methods to detect beaconing with an acceptable FPR of less than 3\% can only be achieved by either using \system, or by applying one of the other two methods with new additional filtering to reduce the FPR after detection.

For MSC detection, \system (AUC=0.85) is the most accurate method; it is followed by the accurate Baywatch-100 (AUC=0.72), Shalginov-WARP (0.58), and the fast Baywatch-10 (0.56). 
Similarly to the beaconing experiment, methods other than \system were unable to produce any detection when configured to a low FPR value of 1\%. 
The lowest possible FPR value for which the two methods produced a detection is as follows: FPR=14.4\% for Baywatch and FPR=3\% for WARP.
In contrast to the beaconing accuracy test, \system dominates the accuracy evaluation for MSC, with higher TPR values for every FPR value, thus making \system more suitable for the detection of MSC in general and for automatic detection in particular.

\textbf{Robustness:} Bots that are configured to periodically communicate with their \cnc server are not always able to do so (e.g., when a compromised device is turned off).
Accordingly, an important aspect to consider when evaluating bot communication detection methods is their ability to detect bots even when some of their routine communication logs are incomplete.
Furthermore, routine communication may be lacking due to an attacker that attempts to evade detection methods by preventing an installed bot from sending some communication.
We refer to this aspect as the \emph{robustness} of a periodicity detection method.
To evaluate the robustness, we designed an experiment that simulates dropped communication and defined a new metric that we refer to as the \emph{robustness score}, which is explained in detail below. 

The experiment for evaluating robustness is similar to the experiment for evaluating accuracy, with two modifications. 
First, we define the notion of a \emph{drop rate} -- the rate at which malicious DNS queries are dropped from the dataset.
This experiment was conducted with 10 different drop rate values ranging from 0\% (i.e., no drop) and increasing by 10\% until a 90\% drop is reached.
Second, we define a acceptable false positive rate that applies for all of the methods, which is crucial for a comparison of the detection rate.
The acceptable false positive rate is determined to be 0.144, because it is the lowest FPR value for which all of the methods are capable of detecting at least one bot in the test set (see Figure~\ref{fig:roc-sssd}).
Eventually, the rate of bots successfully detected while maintaining the specified FPR are reported as the \emph{detection rate}. 

\vspace{-5mm}

\begin{figure*}[h]
\centering
\includegraphics[width=1.0\textwidth]{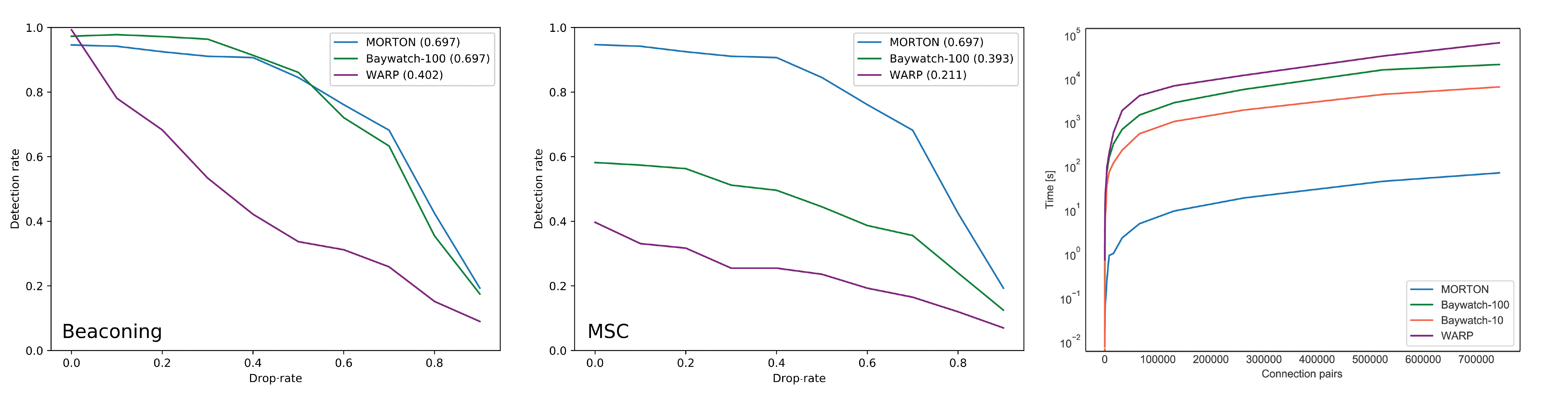}
\caption{The robustness scores of \system, \baywatch, and WARP for beaconing (left) and MSC (middle), and
their run-time (in seconds) with different $m$ values on the test set (right).}
\label{fig:robust_and_time}
\vspace{-2mm}
\end{figure*}

\vspace{-3mm}

The experiment results in a function that maps drop rate values to detection rates, which is henceforth referred to as the \emph{robustness function}.
The area that is captured between the robustness function and the \textit{x}-axis is referred to as the \emph{robustness score}.
The robustness score is a scalar that describes the level of robustness provided by a system that detects periodic signals.
The score ranges between zero and one, and a perfect score of one signifies a system with a perfect detection rate when dealing with varying rates of dropped traffic.
Accordingly, the evaluation of robustness deals with the robustness score of the different methods for both beaconing and MSC.

The robustness scores of all of the methods evaluated are presented in Figure~\ref{fig:robust_and_time}.
For beaconing detection, the highest robustness score (0.697) was achieved by both \system and \baywatch, whereas WARP received a score of 0.402. 
For MSC, \system (0.697) performed best and was followed by \baywatch (0.393) and WARP (0.211).

\textbf{Efficiency:}
To better understand the relationship between the number of devices in the dataset and the run-time performance of the compared methods, we select subsets of the original dataset where the number of devices varies.
This sampling results in 11 subsets of the original datasets with $2^1, 2^2, 2^4, ...  2^{10}$, and $2^{11}$ device IDs that were randomly selected from the original dataset.
The methods are then implemented on each of the datasets sampled to estimate the total number of seconds required for bot communication detection.
The results are portrayed in Figure~\ref{fig:robust_and_time}.
The largest sample consists of 2,048 devices and slightly more than 700,000 CPs. 
For that amount of devices and CPs, \system is the fastest at bot communication detection.
When \system is applied on the entire test set, it takes slightly more than 70 seconds to complete its run. 
The fast \baywatch-10 takes 16 minutes (100x), the accurate \baywatch-100 takes almost three hours (1000x), and WARP takes almost 24 hours (10000x).
The number of devices within a large enterprise network can be significantly larger than 2,048 devices thus resulting in even greater differences.\\
\noindent A summary of the labeled experiment evaluation results is presented in Table~\ref{tab:summary}.
\vspace{-4mm}
\begin{table}[ht] 
\caption{Summary of the results. The best performing method in every category is marked in bold.}
\label{tab:summary} 
\scriptsize
\centering
\begin{tabular}{@{}cccccc@{}}
\toprule
\multicolumn{1}{l}{} Category &
  \multicolumn{1}{l}{} &
  \multicolumn{1}{l}{\system} &
  \multicolumn{1}{l}{Baywatch-10} &
  \multicolumn{1}{l}{Baywatch-100} &
  \multicolumn{1}{l}{WARP} \\ \midrule
AUC                                  & Beaconing            & 0.85            & 0.64             & 0.85           & \textbf{0.97} \\
TPR (FPR=1\%)                         & Beaconing            & \textbf{0.77}   & 0                & 0              & 0             \\
Robustness score                     & Beaconing            & \textbf{0.697}  & \textless{}0.697 & \textbf{0.697} & 0.402         \\
AUC                                  & MSC & \textbf{0.85}   & 0.56             & 0.72           & 0.58          \\
TPR (FPR=1\%)                         & MSC & \textbf{0.77}   & 0                & 0              & 0             \\
Robustness score                     & MSC & \textbf{0.697}  & \textless{}0.393 & 0.393          & 0.211         \\
CP classifications / sec & General              & \textbf{10,000} & 729.1            & 64.8           & 8.1           \\ \bottomrule
\end{tabular}%
\vspace{-2mm}
\end{table}

\vspace{-4mm}

\vspace{-5mm}
\section{Real-World Evaluation} \label{sec:new-real-world}
In the previous section, we compared \system with two proposed methods for malicious periodicity detection on a synthetic labeled dataset. In this section, we provide a complementary analysis, by comparing the methods on an unlabeled DNS dataset, to evaluate its performance in a real-world setting.


\subsection{Methodology}
The methodology of the real-world analysis involves applying the compared methods, namely, \system, \baywatch-100, and WARP on a real-world, unlabeled dataset, and comparing the rates of true positive and false positive detections.
For consistency, the setting matches the description in Section~\ref{sec:eval}. Specifically, the \baywatch setting that optimizes accuracy, namely, \baywatch-100, is compared to analyze the true positive rate in favor of accuracy over run-time.
The dataset for evaluating the real-world efficacy of the methods consists of DNS logs that were produced by two worldwide enterprises consisting of 20,000 users over the course of a week.
There are no labels within the dataset to indicate whether a user is a bot or not.
The results of the detections are either, devices detected by \system or CPs detected by \baywatch-100 or WARP. In order to validate the results, the host names communicated with are analyzed further. Note that unlike CPs, where the host name is an inherent part of the detected pair, the detected devices in \system does not natively consist of a host name. Therefore, for further validation, all of the contacted host names are extracted for every device detected by \system.

In order to validate and label the detections made by the three methods we create two automated labeling techniques.
To validate the maliciousness of the detections, each host name detected is tested against VirusTotal (VT).
A host name is flagged as malicious according to VT if it has at least two detected malicious files communicating to it or if the host itself is flagged malicious by at least two security vendors. Either of these attributes indicates that the host name as known to be used for \cnc communication.
A second validation mechanism is used to validate the periodicity detection regardless of the maliciousness of the detection, namely, is it actually involved in periodic communication (is it a bot). Intervals between consecutive DNS queries are calculated and then counted to validate that the detection is indeed periodic.
Finally, each detection made by either of the three methods is classified as: (1) a bot, which is verified as malicious using VT, (2) a bot, which is unverified as malicious using VT, or (3) not a bot (i.e., a false positive).

\subsection{Evaluation results}
\begin{figure*}
\vspace{-4mm}
\centering
\includegraphics[width=1.0\textwidth]{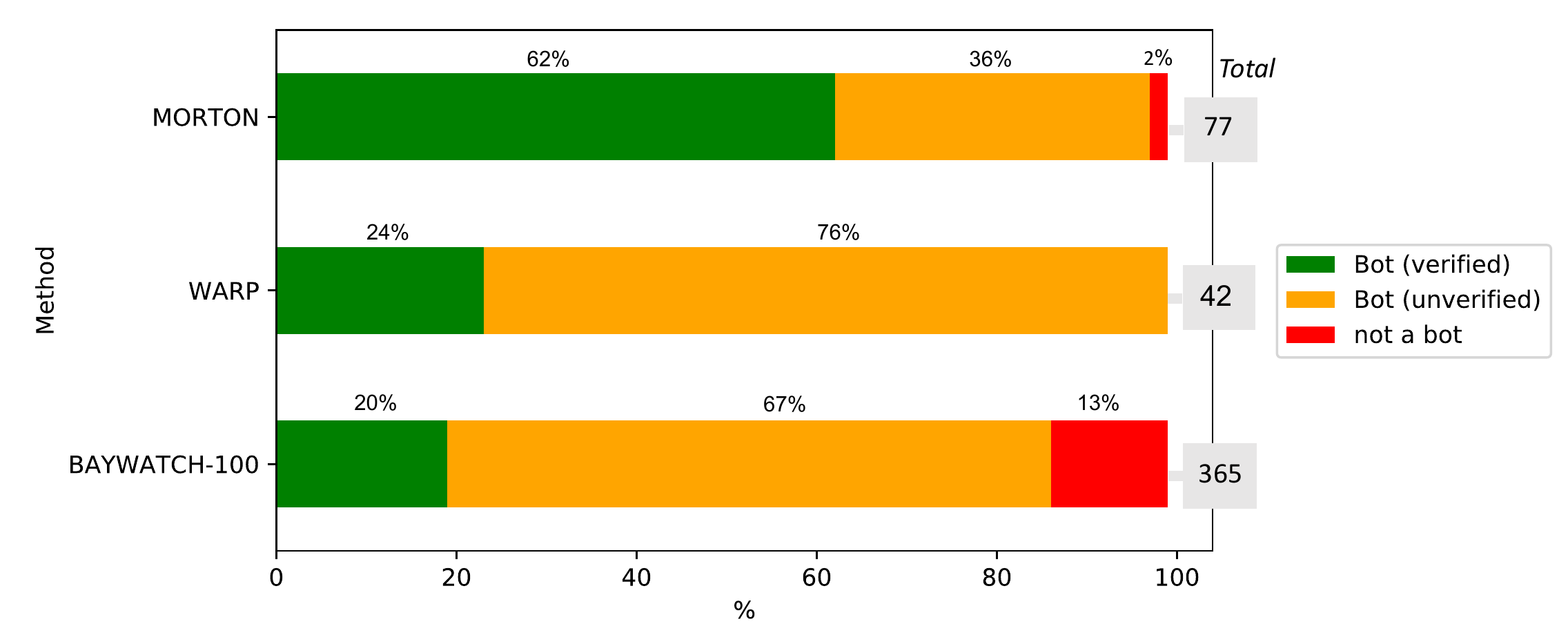}
\caption{A comparison of MORTON, Baywatch-100, and WARP results for unlabeled real-world DNS traffic.}
\label{fig:real_world_comparison}
\end{figure*}
In this experiment, \system is able to detect 77 devices for that week, having 62\% of those devices verified as malicious bots, a figure which is almost three times larger than the true positive rate of both \baywatch-100 and WARP. 
\baywatch-100 detects 365 CPs, which is approximately five times more detections than \system but only 19\% are validated as malicious bots. This drawback is important to note because in a real-world scenario it would result in an excessive amount of time with low return spent by the CERT team analyzing these detections.
WARP achieves the lowest false positive rate (0\%), but this comes with a price of lower coverage, as this method detects only 42 CPs for that week which is nearly half of the detections made by \system with only 2\% false positives.
The real-world experiment results are summarized in Figure~\ref{fig:real_world_comparison}.



\subsection{Case studies}
We analyze three use cases detected during the real-world experiment to understand the nature of the false positives (FPs) and true positives (TPs) made by the compared methods.

\begin{figure}%
\hfill
\subfigure[]{\includegraphics[width=0.36\textwidth]{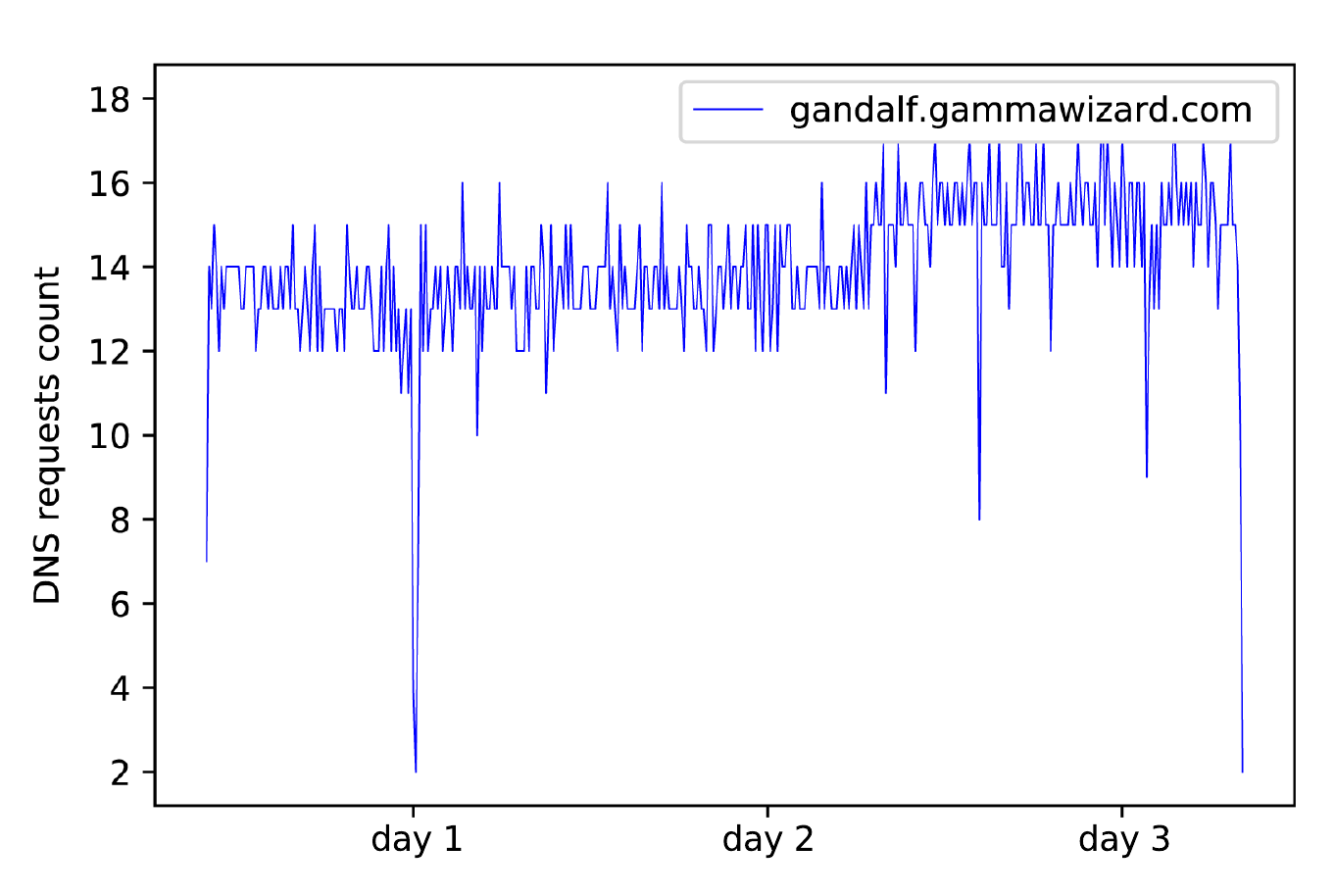}}
\hfill
\subfigure[]{\includegraphics[width=0.36\textwidth]{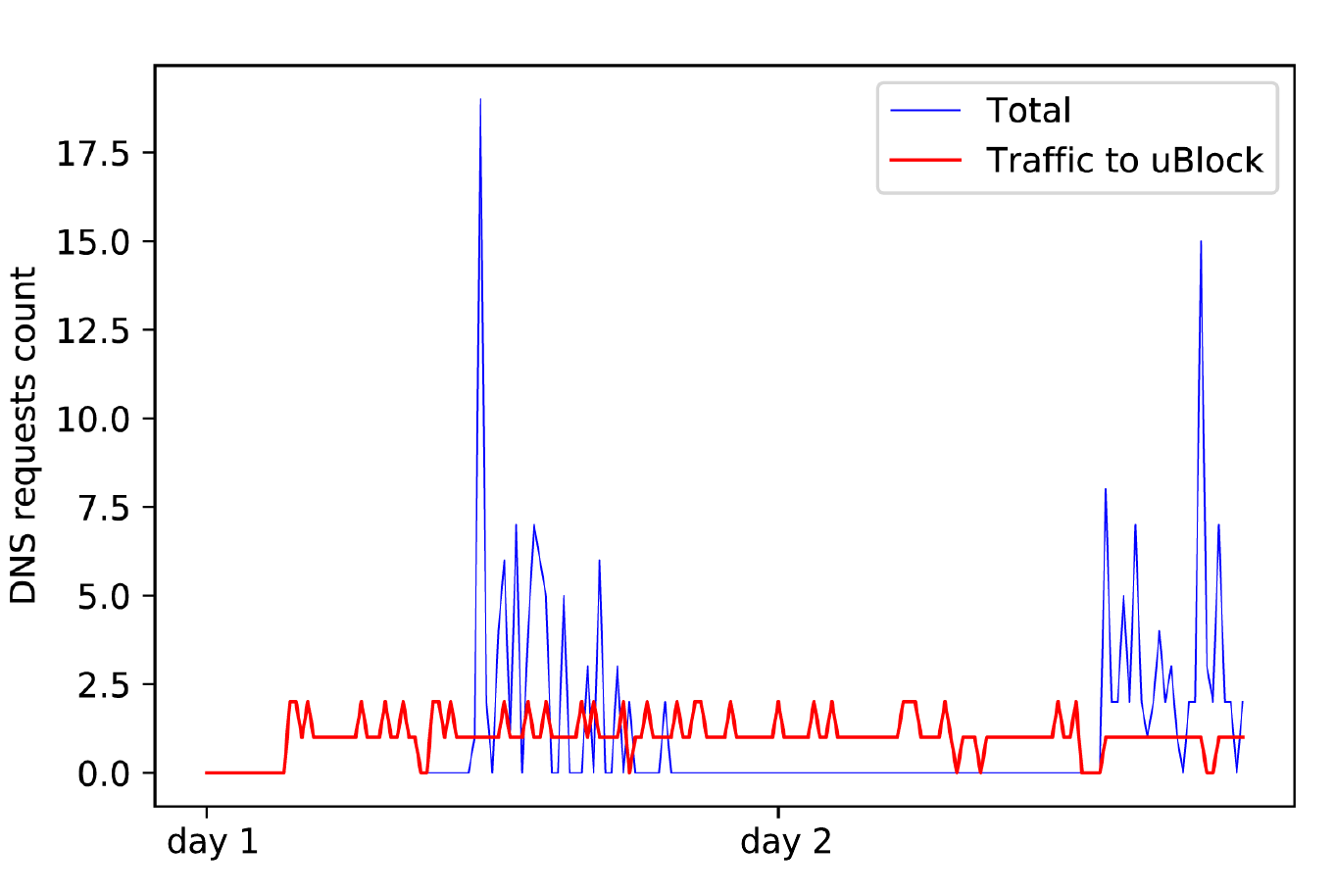}}
\hfill
\subfigure[]{\includegraphics[width=0.36\textwidth]{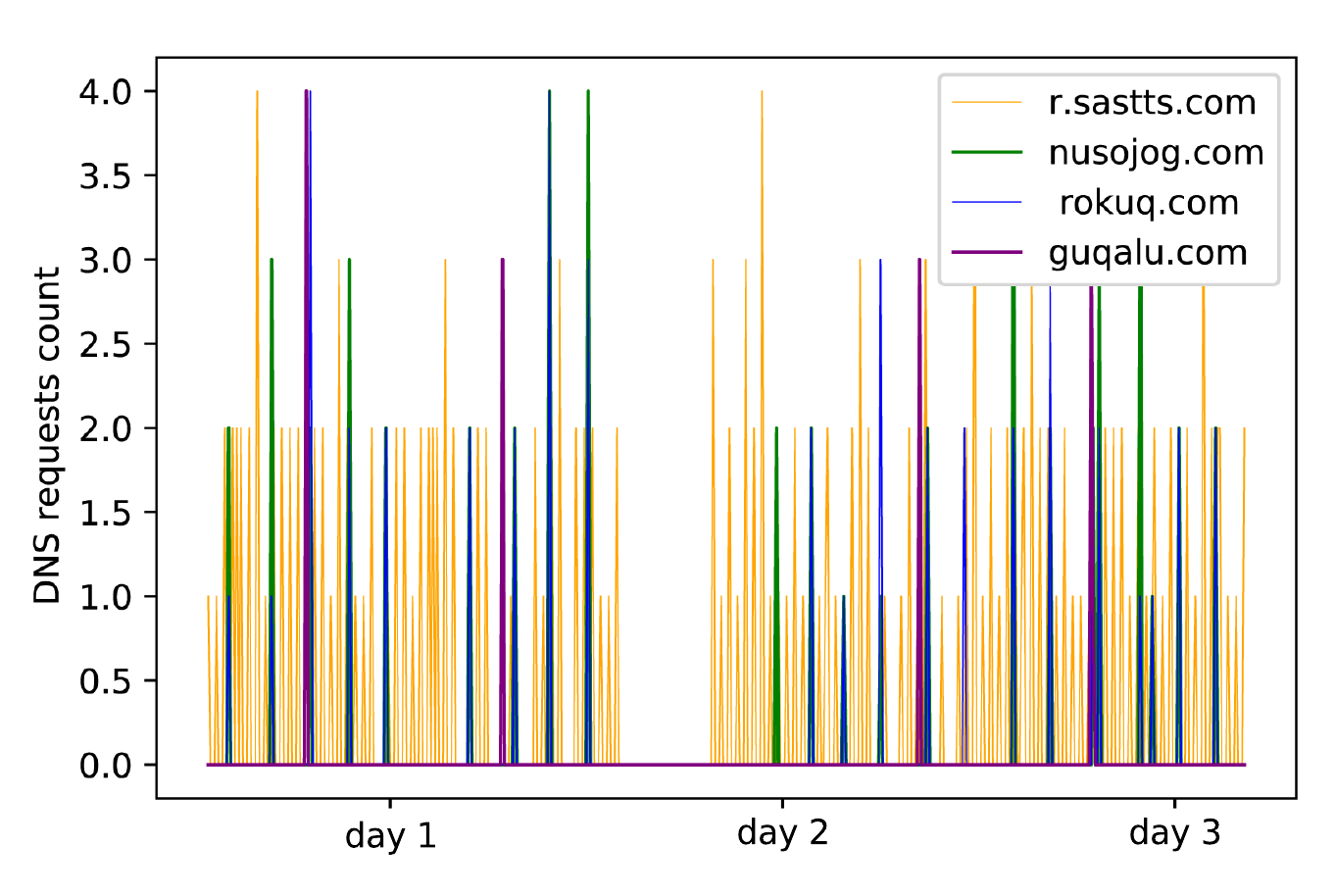}}
\hfill
\caption{Outgoing DNS requests for analyzed threats: (a) "Gandalf" mistakenly detected by \baywatch-100 (b) uBlock uses a single host name technique (c) IsErik adware communicates with its servers using multiple host names.}
\label{fig:realworldexamples}
\end{figure}

\textbf{"Gandalf":} gammawizard.com is a website that provides legitimate online analytics tools for trading. 
\baywatch-100 incorrectly detected one of these tools, called "Gandalf".  
This legitimate activity is in fact periodic (i.e.,  queried every two minutes). However, despite being periodic, the queries made to "Gandalf" does not constitute malware beaconing and consits of a high volume of queries made within a very short time frame, as illustrated in Figure~\ref{fig:realworldexamples}. 
This demonstrates that \system overcomes incorrect detections of this form because it is trained only on periodic queries that match beaconing behavior.

\textbf{uBlock:} \emph{uBlock}~\cite{uBlock} is a malicious add-on for the Chrome browser.
It uses cloned legitimate code to disguise as a legitimate ad blocker while in fact facilitating a malicious backdoor for cookie stuffing, a technique used to commit ad fraud.
uBlock performs DNS queries every 15 minutes to learn the hosting IP address of its servers, to which it sends a ``heartbeat'' messsage (see Figure~\ref{fig:realworldexamples}).
The message is sent as HTTPS GET requests to various URLs that start with ``https://ublockerext.com/heartbeat.''
\system is the only method to identify uBlock, thus demonstrating successful detection of bot communication to a single host name, as shown in Section~\ref{sec:eval}.

\textbf{IsErik:} IsErik~\cite{BitDefenderIsErikReport} is a sophisticated and modular adware~\cite{urban2018towards}, that uses MSC to communicate with its servers.
IsErik's infection process involves the automated and periodic execution of code through WScript on a victim's device. 
The code execution includes communication to a \textit{variety} of host names, as is also documented in public security reports~\cite{TerndMicroIsErikBlog}~\cite{BitDefenderIsErikReport}.
A time series plot of the DNS queries made by IsErik to its \cnc servers is presented in Figure~\ref{fig:realworldexamples}.
The IsErik case study demonstrates the importance of detecting routine communication made to multiple host names --- a feature that distinguishes \system from the compared methods. 

Our interpretation of both the empirical results and our further case studies investigations is that \system is better suited for detecting ~\emph{malicious} periodic communication.
As mentioned in the analyzed cases, this stems from the training process of \system and the use of a supervised classifier, that distinguish \system from the other compared methods. 

To summarize, the experiment results obtained on real-world DNS traffic demonstrate the effectiveness of \system for malicious bot detection in a real-world, unlabeled environment and that \system's true positive rate exceeds that of the two examined proposed methods for bot periodicity detection by as much as three times.
A clear advantage of \system compared to the other methods is the ability to accurately detect \emph{malicious} real-world bot communication, while eliminating legitimate periodic queries, that do not appear within the training set. 
Furthermore, \system also has a proven unique ability to detect bot communication of both a single host name as demonstrated by the uBlock example, and multiple host names as demonstrated in the IsErik use case.

\section{Conclusions} \label{sec:conclusions}
In this paper, we present \system, a system that detects bots by analyzing DNS communication logs and identifying routine DNS queries to disreputable host names that are queried by a bot on behalf of an enterprise device.
\system is designed for large-scale corporate networks, and accordingly, it was evaluated on two large-scale DNS datasets and compared to the recently proposed methods, \baywatch, proposed by Hu \etal~\cite{hu2016baywatch}, and WARP, as proposed by Shalaginov et. al~\cite{shalaginov2016malware}, which were also designed for large-scale networks.
The labeled (supervised) evaluation indicates that \system is comparable to the two methods in terms of accuracy for the detection of malware beaconing but outperforms those methods when detecting bots that communicate using advanced techniques, such as multistage channels.
Additionally, \system was found to be more robust and efficient than the other methods evaluated.
The unlabeled (unsupervised) evaluation on two large enterprises with 20,000 users demonstrated the effectiveness of \system for malicious bot detection in a real-world, unlabeled environment and that \system's true positive rate exceeds that of the other two methods by as much as three times.
Three use cases were analyzed to demonstrate \system's practicality in the real-world setting, as well as to further investigate the difference in performance between the methods. We conclude that \system's design is aimed at the detection of \emph{malicious} periodicity in contrast to the other two methods, thus enabling \system to sustain a high true positive rate, while detecting both single host and MSC bot communication techniques.
\vspace{-4mm}

\bibliographystyle{splncs04}
\bibliography{bibliography}

\begin{subappendices}
\renewcommand{\thesection}{\Alph{section}}

\section{\label{sec:appen1}\vspace{-3mm}Detecting Multiple Host Names}
A bot can communicate with its \cnc server using various techniques involving either a single host name or multiple host names.
Traditionally, bot communication techniques utilize a single host name, because it is easier and less expensive to set up.
Notable examples of such techniques include malware beaconing, local scheduling, scheduled file transfers, connectivity checks, and automated exfiltration~\cite{MITREMatrix,nadler2019detection}.

The primary drawback of bot communication techniques that use a single host name is their lack of robustness.
The single host name is effectively a single point of failure, and if the host name is unavailable for any reason, the attackers cannot control their bots.
An additional drawback is that communication with a single host name may be less covert.
For instance, in DNS data exfiltration, every exfiltration message is sent to an attacker's host name.
A single host name that receives a large volume of exfiltration messages is more detectable by security systems~\cite{nadler2019detection}.
Therefore, the activities of bots that split their DNS exfiltration messages and send them to multiple host names are less suspicious.
The drawbacks of single host techniques are addressed by bot communication techniques that use multiple host names.

The most well-known use of multiple host names for botnet communication is through domain generation algorithms~\cite{MITREMatrix} (DGAs), which are used by over 40 known botnets~\cite{plohmann2016comprehensive}. 
DGAs are used by attackers to generate and select the domain names used to form the \cnc communication channel.
Most bots that use DGAs generate new domain names on a daily basis~\cite{plohmann2016comprehensive}, thus pointing to the importance of detecting bot communication that uses multiple domain names.

Multistage channels (MSC) are another bot communication technique in which multiple host names are used.
The initial installation of the bot on a compromised device is referred to as the first stage of the infection.
Throughout the first stage, the bot communicates with its \cnc through either a single host name or multiple host names.
However, the host names will change when the first stage bot requires an upgrade. 
A bot upgrade typically involves communicating with a new host name to download a module that enhances the bot's capabilities.
The process of upgrading the bot is referred to as the second stage of the infection.
The MSC bot communication technique often involves several stages, where multiple host names are gradually upgrading the bot.
The use of MSC improves the robustness of a botnet's infrastructure, because security researchers cannot easily identify the different host names that will be used by a botnet in order to shut down its operation (i.e., prevent bots from upgrading).

Other cases of bot communication techniques in which multiple host names are used include fallback channels and multihop proxies~\cite{MITREMatrix}.
In fallback channels, a bot that fails to communicate with its \cnc host name attempts to communicate to the host name next in line, based on a prioritized list of host names.
Multihop proxies is a bot communication technique in which the \cnc channel is established through a series of proxy servers that are associated with different host names.
The series of proxy servers between bots and their \cnc servers prevents security researchers from easily matching a bot communicating with its \cnc server based on network logs.
\system is designed to detect every multiple host communication technique mentioned, as long as it is used in a periodic manner.

\begin{figure}[H]
\vspace{-4mm}
\centering
\includegraphics[width=1.0\textwidth]{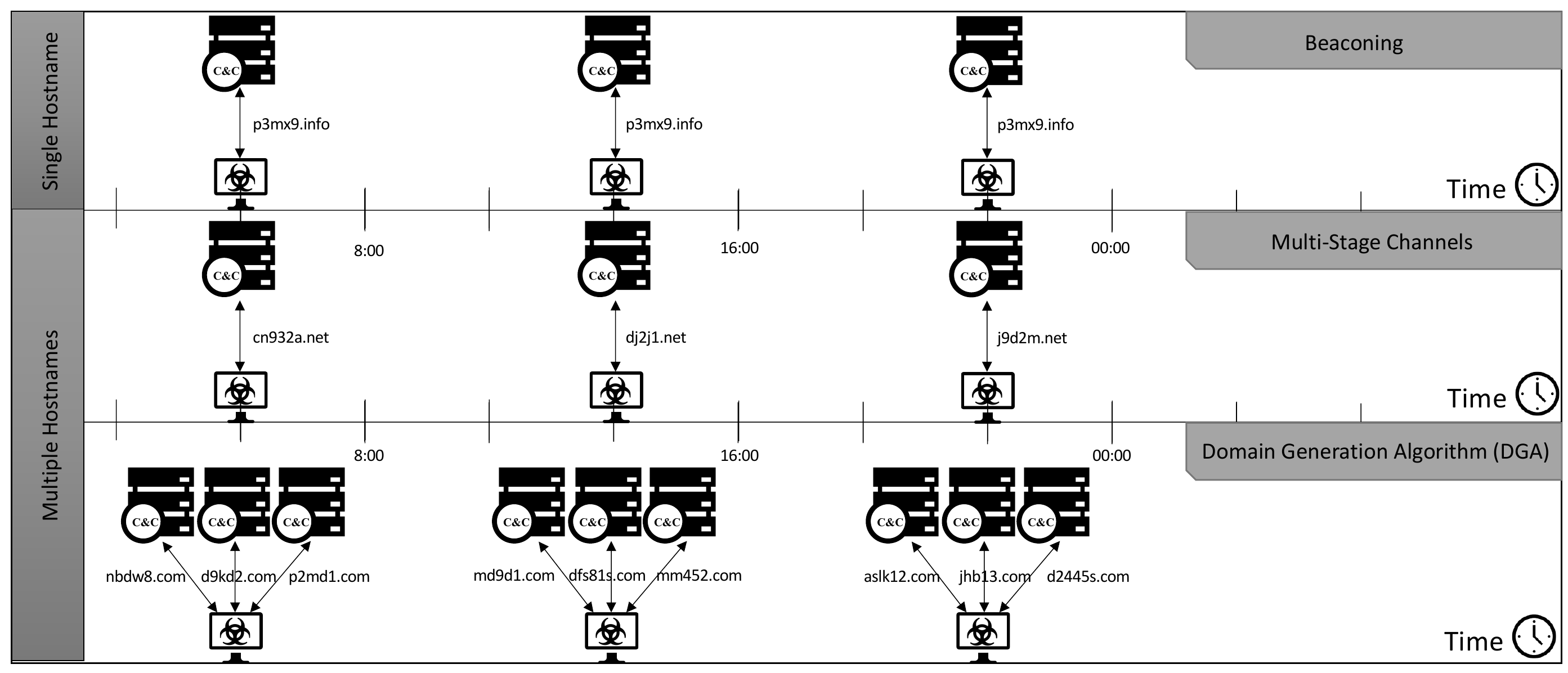}
\caption{Bot communication techniques examples.
An example of a malware beaconing technique with a single host name (top); a MSC technique in which a different host name is used in each stage (middle); and DGA-based \cnc communication where various different host names are used each day (bottom).}
\label{fig:cnc_infrastructure}
\end{figure}

Figure~\ref{fig:cnc_infrastructure} illustrates bot communication techniques that use a single host name or multiple host names.
The first example (top) is of a malware beaconing technique with single host name \cnc communication.
The second example (middle) is of a MSC technique in which a different host name is used in each stage.
The last example (bottom) is of a DGA-based technique where various different host names are used each day for bot communication.

\section{\label{sec:appen2}\vspace{-3mm}Environment}

The entire evaluation is conducted on a cloud computing instance (AWS EC2 C5.x18 xlarge\footnote{https://aws.amazon.com/ec2/instance-types/c5/}), with extensive computational and memory resources, to correctly simulate the hardware of a large-scale enterprise network.
The specification of the instance includes 72 virtual CPUs (vCPU), 144 GB of memory, and a 550GB EBS disk.

\end{subappendices}
	
\end{document}